\documentclass[twocolumn,aps,prl,superscriptaddress
]{revtex4}
\usepackage{amsmath,amsfonts,amssymb,bm,esint,graphicx
}

\newcommand{\Eq}[1]{Eq.~\eqref{#1}}
\newcommand{\eq}[1]{\eqref{#1}}

\newcommand{\beq}{\begin{equation}}
\newcommand{\eeq}{\end{equation}}
\newcommand{\beqa}{\begin{eqnarray}}
\newcommand{\eeqa}{\end{eqnarray}}
\newcommand{\Beqa}{\begin{eqnarray*}}
\newcommand{\Eeqa}{\end{eqnarray*}}

\newcommand{\pdag}{{\phantom{\dagger}}}

\newcommand{\past}{{\phantom{\ast}}}

\newcommand{\pp}{\partial}

\newcommand{\msk}{\mkern 2mu}
\newcommand{\nmsk}{\mkern-2mu}
\newcommand{\smsk}{\mkern 1mu}
\newcommand{\snmsk}{\mkern-1mu}

\newcommand{\triplecolon}{\raisebox{-0.075ex}{\scalebox{0.7}{\bm\vdots}}}

\begin{document}

\title{Viscous Dissipation in One-Dimensional Quantum Liquids}

\author{K. A. Matveev}
\affiliation{Materials Science Division, Argonne National Laboratory, Argonne, Illinois 60439, USA}
\author{M. Pustilnik}
\affiliation{School of Physics, Georgia Institute of Technology, Atlanta, Georgia 30332, USA
}

\begin{abstract}
We develop a theory of viscous dissipation in one-dimensional single-component quantum liquids at low temperatures. Such liquids are characterized by a single viscosity coefficient, the bulk viscosity. We show that for a generic interaction between the constituent particles this viscosity diverges in the zero-temperature limit.  In the special case of integrable models, the viscosity is infinite at any temperature, which can be interpreted as a breakdown of the hydrodynamic description. Our consideration is applicable to all single-component Galilean-invariant one-dimensional quantum liquids, regardless of the statistics of the constituent particles and the interaction strength. 
\end{abstract}

\maketitle

Hydrodynamics is a universal description of fluid flow at long times and distances, when the state of the fluid is completely characterized by only a few macroscopic parameters~\cite{LL_VI}. 
Hydrodynamics ideas have been at the center of both experimental and theoretical attention since the early days of the theory of quantum liquids~\cite{Khalat_book,PN}, originally developed to describe properties of liquid $^4$He~\cite{Landau_SF,Khalat_book,PN} and $^3$He~\cite{Landau_FL,PN,AK_review}. The search for observable manifestations of viscous hydrodynamic flow of electron liquids in solids also has a long history~\cite{Gurzhi}. Although attaining the hydrodynamic regime in this case is difficult because neither the energy nor the momentum of electrons are conserved due to their interaction with crystal lattice and impurities, signatures of hydrodynamic flow were observed in ultraclean two-dimensional~\cite{graphene} and quasi-one-dimensional~\cite{wires} systems.
Viscous hydrodynamic behavior was also observed in three-dimensional~\cite{cold gas_3D} and quasi-one-dimensional~\cite{cold gas_1D} resonantly interacting ultracold Fermi gases, which are free of many complications associated with conventional condensed matter systems.    

Higher-dimensional fluids are characterized by two viscosity coefficients, the shear viscosity $\eta$ arising due to friction between adjacent layers of the fluid, and the bulk viscosity $\zeta$ associated with the expansion or compression of the fluid. At low temperatures, the bulk viscosity is often neglected. For instance, in the Fermi-liquid theory the shear viscosity is proportional~\cite{AK_review,LP_X} to the quasiparticle lifetime, $\eta\propto\tau$, and diverges as $\eta\propto T^{-2}$ at $T\to 0$, whereas \mbox{$\zeta \propto T^4\tau\propto T^2$}~\cite{SB}, which vanishes in the same limit. In one dimension the shear viscosity is obviously absent. This puts focus on the bulk viscosity $\zeta$ and its behavior at low temperatures. 

One-dimensional quantum systems~\cite{Giamarchi} differ dramatically from their higher-dimensional counterparts. Irrespective of the nature of the constituent particles, gapless excitations  can be described in terms of waves of density with acoustic low-energy spectra~\cite{Giamarchi,Haldane_LL}. Similar to fermionic quasiparticles in the Fermi liquid, these excitations are the key ingredients of the Luttinger-liquid theory~\cite{Haldane_LL}. Although this theory has been extremely successful~\cite{Giamarchi}, it is not free of drawbacks~\cite{Andreev,Samokhin, ISG,PKKG}. In particular, it is known~\cite{ISG,Andreev,Samokhin} that the conventional Luttinger-liquid theory is not well equipped to handle phenomena in which relaxation of elementary excitations plays an essential part. Viscous dissipation belongs to this class of phenomena. Below we show that, remarkably, the Fermi-liquid result~$\zeta\propto T^4\tau$ is applicable to both fermionic and bosonic one-dimensional systems. However, because mechanisms of quasiparticle decay in one dimension differ from those in higher dimensions, $\tau$ is anomalously large, resulting in manifestly non-Fermi-liquid temperature dependence: $\zeta$ increases with the decrease of temperature and diverges at $T\to 0$. 

To set the stage, consider a system of $N$ identical particles confined in a one-dimensional interval of length $L$. In the thermodynamic limit when both $N$ and~$L$ are large, precise choice of the boundary conditions is irrelevant. For our purposes, it is convenient to view the system as a ring of circumference~$L$, so that in equilibrium the particle density \mbox{$n = N/L$} is uniform. Imagine now that the size of the system~$L$ slowly changes with time. During such change, 
fluid elements at distance $\Delta x$ move with respect to each other with velocity $\Delta u = (\Delta x/\snmsk L)\smsk\dot L$. (Hereinafter the dot denotes the derivative with respect to time.) This gives rise to the position-independent gradient of the fluid velocity \mbox{$\pp_x u = \dot L/\snmsk L$}. 
Changing the size of the system inevitably leads to an irreversible transfer of energy to it, i.e., heating. The bulk viscosity $\zeta$ is the transport coefficient that controls the heating rate $W\nmsk$. For small~$\pp_x u$, this rate is given by
\beq
W = \zeta L\smsk (\pp_x u)^2.
\label{1}
\eeq  
Below we evaluate the heating rate and then use \Eq{1} to extract the viscosity.

Long-wavelength excitations of gapless one-dimensional systems allow for a continuum description~\cite{Giamarchi,hydrodynamics,Haldane_LL,ISG}, similar to that employed in the theory of superfluidity~\cite{Landau_SF,Khalat_book}. In this approach, the Hamiltonian density of the system is written as an expansion in derivatives of two bosonic fields describing the right- and left-moving excitations. In the leading orders, the right and left movers decouple; hence, it is sufficient to discuss the right movers only. The corresponding part of the Hamiltonian reads~\cite{left}
\beq
H = \frac{\hbar v}{4\pi}\!\int_0^L\! dx\nmsk\left[\nmsk\colon\!(\pp_x\varphi)^2\colon
\nmsk+ \frac{\hbar}{3m_\ast v}\colon\!(\pp_x\varphi)^3\colon\nmsk+\ldots\msk\right]\!,
\label{2}
\eeq
where the field $\varphi$  obeys the commutation relation 
\mbox{$[\pp_x\varphi(x),\varphi(y)] = 2\pi i\smsk\delta(x - y)$} 
and is subject to the periodic boundary condition $\varphi(x) = \varphi(x+L)$; the colons denote normal ordering. The gradient expansion~\eq{2} describes low energy properties of single-component one-dimensional systems with interaction between particles decaying with the distance faster than~$1/x^2$, see, e.g.,~Refs.~\cite{ISG,Pereira,PM_KdV,PM_soliton,RM-14}.  The parameters $v$ and $m_\ast$ in this expansion have the units of velocity and mass, respectively. For Galilean-invariant systems~\cite{Khalat_book,PN,ISG,Pereira,PM_mass} 
\beq
v^2 = \frac{n}{m}\msk\pp_n^2 E_0,
\quad
\frac{m_\past}{m_\ast} = \frac{\smsk\pp_n(vn)}{2v\sqrt{K\smsk}}\smsk,
\label{3}
\eeq
where $E_0$ is the ground state energy per length, $m$ is the mass of the constituent particles, and $K = \pi\hbar\smsk n/m v$ is the Luttinger-liquid parameter~\cite{ISG,Giamarchi}. 
Both $v$ and $m_\ast$ depend on the density and are finite and positive. 

The first term in the expansion \eq{2} describes free bosons with linear spectrum, whereas the second term represents the interaction between these bosons. This term is irrelevant in the renormalization group sense~\cite{Haldane_LL} and can often be neglected, which constitutes the Luttinger-liquid~\cite{Haldane_LL} approximation. It is well-known~\cite{ISG,Samokhin}, however, that naive attempts to account for the interaction perturbatively lead to a divergent inelastic decay rate of the Luttinger liquid bosons. This difficulty is circumvented~\cite{Rozhkov} by rewriting the bosonic Hamiltonian~\eq{2} in terms of effective fermions with the help of the identity~\cite{Giamarchi,Haldane_LL,bosonization} $\psi(x) = L^{-1/2}\colon \!\nmsk e^{i\varphi(x)}\nmsk\colon\nmsk$. In the fermionic representation, the first two terms in the expansion~\eq{2} take the form
\beq
H_0 = \!\int_0^L\!\!dx\smsk\Bigl[-\msk i\hbar v\msk\triplecolon\msk\psi^\dagger(x)\smsk\pp_x\psi(x)\smsk\triplecolon
+ \frac{\hbar^2}{2m_\ast}\msk\triplecolon\msk(\pp_x\psi)^\dagger(\pp_x\psi)\smsk\triplecolon\,
\Bigr],
\label{4}
\eeq
where the symbols $\triplecolon$ denote the normal ordering of fermionic operators with respect to the ground state in which single-particle states with positive (negative) momenta are empty (occupied)~\cite{Haldane_LL,bosonization}. Unlike \Eq{2}, the fermionic Hamiltonian~\eq{4} is quadratic and easily diagonalizable by Fourier transform in the usual way, 
\beq
H_0 = \sum_l \varepsilon_l^\pdag \smsk\triplecolon\,\psi^\dagger_l\psi^\pdag_l\triplecolon\,,
\quad
\varepsilon_l = vp_l + \frac{p_l^2}{\smsk 2m_\ast}.
\label{5}
\eeq
Here \mbox{$\psi_l = L^{-1/2}\!\int_0^L\!dx\,\psi(x)\smsk e^{-ip_l x/\hbar}$} and the single-particle momenta are given by \mbox{$p_l = (2\pi\hbar\smsk/L)\smsk l$} with integer $l$. In the fermionic language, the periodic boundary condition on~$\varphi$ translates to the constraint \mbox{$\int_0^L\!dx\,\triplecolon\msk\psi^\dagger(x)\psi(x)\smsk\triplecolon\msk = 0$}, which can be also written as
\beq
\sum_l  \msk\triplecolon\,\psi^\dagger_l\psi^\pdag_l\triplecolon\, = 0\msk.
\label{6}
\eeq

The fermions described by Eqs.~\eq{4}-\eq{6} emerged as a result of the exact diagonalization of the first two terms in the gradient expansion~\eq{2}. These fermions do not represent exact eigenstates of the full Hamiltonian, but should be viewed instead as fermionic quasiparticles, akin to those in the Fermi-liquid theory~\cite{PN,AK_review,Landau_FL}. Interaction between the quasiparticles originates in higher-order contributions in the gradient expansion~\eq{2}. Whereas in the conventional Fermi-liquid theory~\cite{PN,Landau_FL,AK_review} such interaction affects the excitation spectrum, in our case the interaction leaves the first two terms in the expansion of $\varepsilon_l$ given in \Eq{5} intact~\cite{left,Rozhkov,PM_KdV,PM_mass}. It is crucial, however, that in the absence of integrability~\cite{Sutherland_book} the quasiparticles acquire a finite decay rate~\cite{ISG,PM_soliton,LFG-07,KPKG-07,PWA-09,MF-13,RM-13,ABG-14,RM-14,PGM-14}.

Similar to the Fermi-liquid theory, the system can be described by the distribution function \mbox{$f_l = \langle\psi^\dagger_l\psi^\pdag_l\rangle$}. Its equilibrium form $f_{0l}$ maximizes the entropy 
\mbox{$S\smsk[f] = -\msk\sum_l\bigl[f_l\ln\snmsk f_l + (1-f_l)\ln\smsk(1-f_l)\bigr]$}
at constant energy and number of quasiparticles. This immediately leads to the Fermi-Dirac distribution
\beq
f_{0l} = \frac{1}{e^{\msk\beta\smsk(\varepsilon_l - \smsk\mu)} + 1},
\label{7}
\eeq 
where $\beta = 1/T$ and $\mu$ is the chemical potential determined by the condition 
\mbox{$\sum_l \bigl[\smsk f_{0l} - f_{0l}\bigr|_{T = 0}\smsk\bigr] = 0$} that follows from \Eq{6}. 
In the leading order in temperature this condition yields 
$\mu = (\pi^2\nmsk/6)\smsk (T^2\nmsk/m_\ast v^2)$.

If the system size depends on time, $L = L(t)$, both $\beta$ and $\mu$ in \Eq{7} are time dependent, as is the excitation spectrum $\varepsilon_l$. Indeed, the velocity $v$ and the mass~$m_\ast$ in \Eq{5} depend on the density $n$ that scales with the size as $1/L$, and the momenta $p_l$ scale the same way. Taking into account the relation  \mbox{$\pp_x u = \dot L/\snmsk L$}, we obtain the equations
\beq
\dot n = - \smsk n\smsk\pp_x u,
\quad
\dot p_l = - \smsk p_l\msk\pp_x u.
\label{8}
\eeq
(In the first equation here one may recognize the continuity equation in the special case \mbox{$\pp_x n = 0$}.) 

When the system size changes, collisions between the quasiparticles cause a change of the occupation numbers~$f_l$. Because collisions preserve the number of quasiparticles and the energy, the time-dependent distribution function obeys the conservation laws~\cite{LP_X} 
\beq
\sum_l\dot f_l = 0,
\quad
\sum_l\varepsilon_l\dot f_l = 0\smsk.
\label{9}
\eeq
If the size changes slowly, $f_l$ remains close to equilibrium at all times. For $f_l = f_{0l}$ Eqs.~\eq{7} and \eq{9} imply that $\dot S[f_0] =\sum_l \dot f_{0l}\ln(1/\snmsk f_{0l} - 1) = 0$, which corresponds to an adiabatic process. Using Eqs.~\eq{7}-\eq{9} and the expression for $\mu$ given above, we find 
\beq
\dot\beta = \beta\bigl[n\msk\pp_n\nmsk\ln(nv)\bigr]\pp_x u,
\quad
\dot\mu = 
\mu\nmsk\left[n\msk\pp_n\nmsk\ln\!\left(\frac{m_\ast}{n^2}\nmsk\right)\snmsk\right]
\nmsk\pp_x u
\label{10}
\eeq
in the leading order in $T/m_\ast v^2 \ll 1$. Differentiation of \Eq{7} with the help of these relations yields
\beq
\dot f_{0l}
= f_{0l}\bigl(1 - f_{0l}\bigr)\smsk
\frac{\smsk 3\smsk v^2p_l^2- \pi^2T^2}{6 m_\ast v^2 T\,}
\smsk\bigl(n\msk\pp_n\nmsk\ln\nu\bigr)\smsk\pp_x u
\label{11}
\eeq
with
\beq
\nu = \frac{\pi\hbar n}{m_\ast v} 
= \sqrt{K\smsk}\bigl(1 - n\msk\pp_n\nmsk\ln\nmsk\sqrt{K\smsk}\smsk\bigr),
\label{12}
\eeq
where we took into account \Eq{3}.

The dependence on time of the full distribution function \mbox{$f_l = f_{0l} + \delta \nmsk f_l$} is governed by the equation
\beq
\dot f_l = I[\smsk f\smsk],
\label{13}
\eeq 
where the functional $I[f]$ is the collision integral~\cite{LP_X}. The detailed balance~\cite{LP_X} ensures that $I[f_0] = 0$ for any Fermi-Dirac distribution, including that given by \Eq{7}. Therefore, whereas in the left-hand side of \Eq{13} the nonadiabatic correction $\delta\nmsk f_l$ can be neglected, it must be retained in the right-hand side. 

Solutions of \Eq{13} satisfy the conservation laws~\eq{9}. Since $f_{0l}$ obeys these laws, so does $\delta\nmsk f_l = f_l - f_{0l}$, which yields the relation \mbox{$\sum_l(\varepsilon_l - \mu)\smsk\delta\snmsk\dot f_l = 0$}. With this relation taken into account, the heating rate $W = T\dot S$ in the lowest nonvanishing order in $\delta\snmsk f_l$ assumes the form
\beq
W = -\smsk T\sum_l\frac{\dot f_{0l}\msk\delta\snmsk f_{l}}{f_{0l}\bigl(1 - f_{0l}\bigr)}.
\label{14}
\eeq 
In order to estimate the viscosity, we use the relaxation time approximation for the collision integral, 
\mbox{$I[f] = -\msk\delta\snmsk f_l/\tau$}, which leads to \mbox{$\delta\snmsk f_l = -\msk\tau\snmsk\dot f_{0l}$}. Equations~\eq{11} and \eq{14} then yield \Eq{1} with
 \beq
\zeta = \mathcal A\msk\frac{\nu^2 n\msk}{\msk(\hbar nv)^3}\msk T^4\smsk\tau\msk,
\quad
\mathcal A = c\smsk\bigl(n\msk\pp_n\nmsk\ln\nu\bigr)^2, 
\label{15}
\eeq
where $c$ is a numerical coefficient of order unity. (The accuracy of the relaxation time approximation is insufficient to allow its evaluation.) Equation~\eq{15} is the main result of our paper. It relates the bulk viscosity $\zeta$ to the lifetime $\tau$ of fermionic quasiparticles. 
In the remainder of the paper we employ \Eq{15} to estimate the viscosity in several typical situations. 

The decay rate $1/\tau$ is determined by higher-order terms in the gradient expansion~\eq{2}, which in turn depend on the interaction potential in the underlying microscopic model. 
For short-range potentials that fall off with the distance faster than any power, the dominant contribution to the rate comes from scattering processes involving three quasiparticles that do not move in the same direction~\cite{ISG,LFG-07,KPKG-07,PWA-09,MF-13,RM-14,ABG-14,PGM-14}. At quasiparticle energies of order $T$, this contribution has the form
\beq 
\frac{1}{\tau} = \mathcal B\smsk\nu^2\smsk nv\!\left(\nmsk\frac{T}{\hbar n v}\nmsk\right)^{\!7},
\label{16}
\eeq 
whereas collisions between three quasiparticles moving in the same direction yield $1/\tau \propto T^{14}$~\cite{PGM-14}. The dimensionless coefficient $\mathcal B$ in \Eq{16} depends on the interaction potential and can be related~\cite{MF-13} to the cubic term in the expansion of the excitation spectrum $\varepsilon_l$. Equations~\eq{15} and~\eq{16} yield 
\beq
\zeta = \frac{\mathcal A}{\mathcal B}
\nmsk\left(\nmsk\frac{\hbar n v}{T}\nmsk\right)^{\!3}\!\hbar n,
\label{17}
\eeq
which diverges at $T\to 0\smsk$. 

It is instructive to apply \Eq{17} to weakly interacting fermions and bosons. We start with spinless fermions. In the absence of interactions $m_\ast = m$ and $v$ coincides with the Fermi velocity~$v_F = \pi\hbar\smsk n/m$, hence $\nu = 1$ [see \Eq{12}], and \Eq{15} gives $\mathcal A = 0$. This property can be traced back to the excitation spectrum, which in this case is given by $\varepsilon_l = v_F p_l \smsk+\smsk p_l^2/2m$, so that $\varepsilon_l$ scales as~$1/L^2$ with the system size. Therefore, the occupation numbers given by the Fermi-Dirac distribution~\eq{7} with both $T$ and~$\mu$ rescaled as $1/L^2$ are independent of time for any~$L(t)$, in agreement with \Eq{11} for $\nu = 1$. Importantly, the relation $\mathcal A = 0$ does not imply vanishing viscosity. Indeed, for free fermions there is no quasiparticle decay; hence, $\mathcal B$ in \Eq{16} also vanishes, rendering \Eq{17} inapplicable. 

The uncertainty is resolved by considering the limit of vanishing (as opposed to neglected from the outset) interaction between fermions. Correction to $\nu$ appears in the first order of perturbation theory in the interaction strength $V(x)$, which gives $\mathcal A\propto V^2$ [see \Eq{15}]. The coefficient $\mathcal B$ in \Eq{16} scales as the probability of three-particle scattering. In perturbation theory, the nonvanishing amplitude of such scattering arises in the second order~\cite{KPKG-07}, hence $\mathcal B\propto V^4$. Therefore, the ratio~$\mathcal A/\mathcal B$ in \Eq{17} scales as the inverse square of the interaction strength and diverges when the interaction is taken to zero, indicating a breakdown of the hydrodynamic description. Consider, for example, spinless fermions with interaction potential $V(x)$ that falls off rapidly with $|x|$ at \mbox{$|x|\gtrsim a_0$} and is smooth and featureless at \mbox{$|x|\lesssim a_0$} (cf. Refs.~\cite{ISG,LFG-07,KPKG-07,PGM-14}).  For $a_0 n\gg 1$ we find $\mathcal A \sim (V_0/\hbar v_F)^2$ and~\cite{KPKG-07,PGM-14} \mbox{$\mathcal B\sim (V_0/\hbar v_F)^4 (a_0 n)^4$} with $V_0 = \int\!dx\msk V(x)\ll \hbar v_F$. Accordingly, $\mathcal A\smsk/\mathcal B$ in \Eq{17} indeed diverges in the free-fermion limit $V_0/\hbar v_F\to 0$.

As an example of a system of bosons with short-range repulsion, we consider bosonic atoms confined in a one-dimensional trap~\cite{BDZ,PS}. Relevant microscopic parameters in this case are the $s$-wave scattering length in three dimensions~$a$ and the amplitude of zero-point motion in transverse direction~$b$~\cite{PS,Olshanii,bosons}. In the first order in interaction, the ground state energy can be estimated as~\cite{PS} $E\sim \hbar^2 N^2 a/mV_{\smsk\text{3D}}$, where $V_{\smsk\text{3D}}$ is the three-dimensional volume occupied by the bosons. For $V_{\smsk\text{3D}}\sim b^2\snmsk L$ this gives \mbox{$E_0 = E/L\sim (\hbar^2\nmsk/m)(an^2\nmsk/b^2)$} for the ground state energy per length. Equation \eq{3} then yields \mbox{$K\sim \sqrt{b^2n/a\smsk}\,$} for the Luttinger-liquid parameter. (For weak repulsion considered here $K\gg 1$.) Substituting this expression into Eqs.~\eq{12} and~\eq{15}, we find $\mathcal A\sim 1$. Because the dimensionless amplitude of three-particle scattering is of order $(a/b)^2$~\cite{bosons}, the decay rate of fermionic quasiparticles evaluated in Refs.~\cite{RM-14,ABG-14} assumes the form of \Eq{16} with \mbox{$\mathcal B\sim (bn)^4$}. The viscosity is then given by \Eq{17} with $\mathcal A/\mathcal B\sim (bn)^{-4}$. 

Interestingly, the viscosity diverges in the limit $bn\to 0$ taken with both $K$ and $\hbar nv/\smsk T$ kept constant. This limit corresponds to the Lieb-Liniger model~\cite{Lieb,Sutherland_book} describing bosons with contact repulsion, which is well known to be integrable. In fact, divergent viscosity is a general property of integrable models. Indeed, in all such models the fermionic quasiparticles introduced in Eqs.~\eq{4} and~\eq{5} represent exact eigenstates; hence, their decay rate vanishes identically~\cite{ISG,LFG-07,KPKG-07,MF-13,RM-14,ABG-14,PM_soliton}. Because $\mathcal A$ is finite for any nonvanishing interaction regardless of integrability, this observation and \Eq{17} imply that $\zeta$ is infinite even at finite temperature.

Equations~\eq{16} and \eq{17} are applicable only when interactions decay rapidly with the distance between particles. An important exception is the system of spinless fermions with interaction potential that falls off with the distance as~$1/|x|^3$. Such interaction describes electrons in quantum wires~\cite{RM-13,wire} and atoms in ultracold dipolar gases~\cite{dipolar gas}.  Electrons, for instance, interact via the Coulomb potential screened by a metal gate at distance~$d$ from the wire. For simplicity, we assume that $d \sim 1/n$. Correction to the velocity then reads $\delta v \sim e^2\nmsk/\hbar\msk$. Substituting $K = 1- \delta v/v_F$ into Eqs.~\eq{12} and \eq{15}, we find \mbox{$\mathcal A\sim (n a_B)^{-2}$}, where \mbox{$a_B = \hbar^2\nmsk/m e^2$} is the Bohr radius. Because of the long-range nature of the potential $V(x)\propto |x|^{-3}$, the quasiparticle decay rate due to scattering of three electrons moving in the same direction scales as $1/\tau\propto T^6$~\cite{RM-13} as opposed to $1/\tau\propto T^{14}$~\cite{PGM-14} for rapidly decaying potentials.  At low temperatures, this contribution dominates the decay rate. Using the result of Ref.~\cite{RM-13} and \Eq{15}, we obtain
\beq
\zeta\sim\frac{\msk\hbar n \smsk(n a_B)^2}{\ln^2\snmsk(m v^2_F/\smsk T)}
\nmsk\left(\snmsk\frac{m v^2_F}{T\msk}\snmsk\right)^{\!2}
\label{18}
\eeq
for the viscosity. For the same reason as for fermions with short-range interaction, $\zeta$ given by \Eq{18} diverges in the free-fermion limit $n a_B\to \infty$. However, the temperature dependence is different from that in \Eq{17}.

We now discuss potential applications of our theory. Our main result, \Eq{15}, relates the viscosity $\zeta$ to the quasiparticle lifetime $\tau$. In principle, the lifetime can be measured in momentum-resolved tunneling experiments~\cite{Yacoby-10}. However, this method is effective only for high-energy excitations with decay rate $\hbar/\tau\gtrsim T$. On the other hand, $\tau$ in \Eq{15} is the lifetime of quasiparticles with energy of order $T$ for which $\hbar/\tau \ll T$. Thus, our theory relates the decay rate of such thermal excitations evaluated in Refs.~\cite{ISG,RM-13,ABG-14,PGM-14} to an experimentally accessible quantity, the viscosity.

Bulk viscosity $\zeta$ can be measured in transport experiments with quantum wires. Indeed, when electron liquid flows past a very smooth obstacle its density changes. The resulting viscous dissipation yields a contribution to the electrical resistance of the wire proportional to $\zeta$~\cite{Vignale,AKS,DM-15}.  Alternatively, viscous dissipation can be measured in cold atomic gases in one-dimensional traps~\cite{BDZ}. In this case, slow changes of longitudinal confinement will heat the gas, see \Eq{1}. This heating can be measured in time-of-flight experiments~\cite{BDZ}.

Although our consideration relied on Galilean invariance, some aspects of our theory are applicable to chiral one-dimensional systems as well. Indeed, the gradient expansion~\eq{2} coincides~\cite{PGM-14} with the well-known~\cite{Wen} effective low-energy description of gapless edge excitations in quantum Hall systems. Consider, for example, a quantum Hall antidot, i.e., a gate-defined depleted region in a two-dimensional electron gas in the quantum Hall regime~\cite{antidot}. Because bulk degrees of freedom are gapped, a low-frequency signal applied to the gate that controls the size of the antidot will pump energy to propagating states at its edge. The resulting temperature rise may be detected by transport spectroscopy~\cite{antidot}. Chiral edge excitations carry charge and interact via Coulomb potential screened by the gate. Our theory then suggests that the energy transfer rate depends on temperature according to $W \propto (T\ln T)^{-2}$, see Eqs.~\eq{1} and \eq{18}. 

\begin{acknowledgments} 

We thank A. V. Andreev and W. DeGottardi for numerous discussions. 
This work was supported by the US Department of Energy, Office of Science, Materials Sciences and Engineering Division. We are grateful to the Aspen Center for Physics (NSF Grant No. PHYS-1066293) for hospitality.
\end{acknowledgments}

\vspace{-3ex}
\onecolumngrid


\begin{thebibliography}{99}
\bibitem{left}
The remainder of the gradient expansion \eq{2} contains terms mixing the bosonic fields $\varphi_R$ and $\varphi_L$ describing, respectively, the right- and left-moving excitations. [In \Eq{2}, $\varphi = \varphi_R$.] The leading contributions of this type are integrals of $(\pp_x \varphi_R)^2(\pp_x\varphi_L)$ and $(\pp_x \varphi_L)^2(\pp_x\varphi_R)$. As shown in Ref.~\cite{Rozhkov}, these perturbations do not alter the first two terms in the expansion of the excitation spectrum given in \Eq{5}. On the other hand, accounting for these perturbations is crucial~~\cite{ISG,LFG-07,KPKG-07,PWA-09,MF-13,RM-14,ABG-14,PGM-14} for the evaluation of the decay rate~\eq{16}.
\end{thebibliography}
\end{document}